\documentclass[prl,twocolumn,superscriptaddress,amsmath,amssymb,aps,longbibliography,nofootinbib,10pt,floatfix,raggedbottom]{revtex4-2}

\usepackage[T1]{fontenc}
\usepackage[utf8]{inputenc}
\usepackage{amsmath,amssymb,amsfonts,bm}
\usepackage{wasysym}
\usepackage{graphicx}
\usepackage{txfonts}
\usepackage[dvipsnames]{xcolor}
\usepackage[colorlinks=true,urlcolor=MidnightBlue,citecolor=MidnightBlue,linkcolor=MidnightBlue]{hyperref}
\usepackage[capitalise]{cleveref}
\definecolor{darkblue}{RGB}{46,49,146}
\hypersetup{linkcolor=darkblue,anchorcolor=darkblue,citecolor=darkblue,
  pdftitle={Topological and Trivial Valence-Bond Orders in Higher-Spin Kitaev Models},
  pdfauthor={Xing-Yu Zhang, Qi Yang, Philippe Corboz, Jutho Haegeman, Yuchi He}}
\crefname{figure}{Fig.}{Figs.}
\Crefname{figure}{Figure}{Figures}

\makeatletter
\let\arxiv@hangfrom@section\@hangfrom@section
\let\arxiv@sectioncntformat\@sectioncntformat
\newcommand{\startSupplement}{%
  \clearpage
  \let\@hangfrom@section\arxiv@hangfrom@section
  \let\@sectioncntformat\arxiv@sectioncntformat
  \setcounter{section}{0}%
  \setcounter{subsection}{0}%
  \setcounter{figure}{0}%
  \setcounter{table}{0}%
  \setcounter{equation}{0}%
  \@removefromreset{equation}{section}%
  \renewcommand{\thesection}{S\arabic{section}}%
  \renewcommand{\thefigure}{S\arabic{figure}}%
  \renewcommand{\thetable}{S\arabic{table}}%
  \renewcommand{\theequation}{S\arabic{equation}}%
  \def\theHsection{supplement.\arabic{section}}%
  \def\theHfigure{supplement.\arabic{figure}}%
  \def\theHtable{supplement.\arabic{table}}%
  \def\theHequation{supplement.\arabic{equation}}%
}
\makeatother

\begin{document}

\title{Topological and Trivial Valence-Bond Orders in Higher-Spin Kitaev Models}

\author{Xing-Yu Zhang}
\thanks{These authors contributed equally to this work.}
\affiliation{Department of Physics and Astronomy, Ghent University, Krijgslaan 281, 9000 Gent, Belgium}

\author{Qi Yang}
\thanks{These authors contributed equally to this work.}
\affiliation{Institute for Theoretical Physics Amsterdam and Delta Institute for Theoretical
Physics, University of Amsterdam, Science Park 904, 1098 XH Amsterdam, The
Netherlands}

\author{Philippe Corboz}
\affiliation{Institute for Theoretical Physics Amsterdam and Delta Institute for Theoretical
Physics, University of Amsterdam, Science Park 904, 1098 XH Amsterdam, The
Netherlands}
\author{Jutho Haegeman}
\affiliation{Department of Physics and Astronomy, Ghent University, Krijgslaan 281, 9000 Gent, Belgium}

\author{Yuchi He}
\email{yuchi.he@ugent.be}
\affiliation{Department of Physics and Astronomy, Ghent University, Krijgslaan 281, 9000 Gent, Belgium}
\affiliation{Rudolf Peierls Centre for Theoretical Physics, Clarendon Laboratory, Parks Road, Oxford OX1 3PU, United Kingdom}

\def\thefootnote{*}
\def\thefootnote{\arabic{footnote}}  

\begin{abstract}
    We investigate the quantum phases of higher-spin Kitaev models using tensor network methods. Our results reveal distinct valence-bond-ordered phases for spin-1, spin-$\tfrac{3}{2}$, and spin-2 models.  In all cases, we find translational symmetry breaking with unit cells being tripled by forming valence-bond orders. However, these three phases are distinct, forming plaquette order, topological dimer order, and nontopological dimer order, respectively. Our findings are based on a cross-validation between variational two-dimensional tensor network calculations: an unrestricted exploration of symmetry-broken states versus the detection of symmetry breaking from cat-state behavior in symmetry-restricted states. The origin of different orders can also be understood from a theoretical analysis.
    Our Letter sheds light upon the interplay between topological and symmetry-breaking orders as well as their detection via tensor networks. 
\end{abstract}

\makeatletter
\@author@finish
\let\arxiv@AAC@list\@AAC@list
\let\arxiv@AFF@list\@AFF@list
\let\arxiv@AFG@list\@AFG@list
\makeatother
\maketitle

For quantum spin lattice systems, valence-bond solids (VBS) break translational symmetry by forming valence-bond orders: some bonds spontaneously become stronger with lower energies than others.  A quantum spin liquid (QSL) was originally defined by the absence of valence-bond orders and magnetic orders~\cite{anderson1973resonating}.  The competition between QSL and VBS is determined by whether tunneling and resonance processes remain dominant in the thermodynamic limit.  Historically, this competition echoes concepts from chemistry (e.g., Kekul\'e bond resonance) and solid-state physics~\cite{pauling1933nature,peierls1955quantum}.  After the initial proposal of QSLs, revolutionary concepts such as emergent gauge fields~\cite{PhysRevB.37.3774}, anyons~\cite{PhysRevLett.59.2095}, topological orders~\cite{Wen}, and long-range entanglement~\cite{chen2010local} were introduced to characterize general QSLs. These characterizations reshape the definition of a QSL and motivate a renewed examination of the interplay between valence-bond order and QSL concepts.

The Kitaev-type spin-spin interaction~\cite{KITAEV20062,PhysRevLett.102.017205} involves nontrivial energetics of quantum spin orientation and dimer resonance, providing a platform to study the interplay of valence-bond order, gauge theory, and topological order.  For comparison, the antiferromagnetic Heisenberg interaction ($J \mathbf{S}_i \cdot \mathbf{S}_j, J>0$) has the singlet pairing as its lowest energy state, thereby giving rise to the well-known dimer picture and motivating dimer models based on dimer resonance~\cite{PhysRevLett.61.2376, PhysRevB.64.144416, PhysRevLett.115.217202}.  This picture can capture various conventional VBS, e.g., Refs.~\cite{PhysRevLett.113.027201,PhysRevLett.110.127203,PhysRevB.84.024406,PhysRevB.88.165138,PhysRevB.91.060403,PhysRevB.91.100407,PhysRevB.87.115144,Zayed2017,science.adc9487}  which lack QSL characters. By contrast, the Kitaev type of spin-spin interaction per bond is Ising-like (e.g., $K_x S_i^x S_j^x$) with two degenerate spin orientations with lowest energy.  Therefore, the possible valence-bond orders of Kitaev models are not captured by minimal dimer models.  The spin-$\tfrac{1}{2}$ Kitaev honeycomb model~\cite{KITAEV20062}, an exactly solvable QSL with a $\mathbb{Z}_2$ gauge field, lacks valence-bond order.  Likewise, the higher-spin Kitaev models also exhibit local $\mathbb{Z}_2$ fluxes as conserved quantities~\cite{PhysRevB.78.115116}, which preclude any magnetic order. Analytical constructions for emergent fermions and braiding~\cite{PhysRevB.78.115116,PhysRevResearch.2.022047, PhysRevB.105.L060403,PhysRevLett.130.156701} have been found for half-integer spin $S$ but not for integer spin.  Semiclassical analyses near the large-spin limit suggest the possibility of valence-bond order with a tripled unit cell~\cite{PhysRevB.78.115116,PhysRevE.82.031113,Rousochatzakis2018}.  Furthermore, the possibility of a spontaneous valence-bond order coexisting with topological order has been proposed~\cite{Rousochatzakis2018}.  

Significant computational efforts~\cite{JPSJ.87.063703,PhysRevB.99.104408,minakawa2020magnetic,PhysRevB.102.121102,PhysRevResearch.2.022047,PhysRevResearch.2.033318,PhysRevB.105.L060403,PhysRevResearch.3.013160,Jin2022,PhysRevB.108.075111,Jahromi2024,PhysRevB.110.035121} have already been devoted to the spin-1 and spin-$\tfrac{3}{2}$ Kitaev honeycomb model.  Such reference points at or near the pure Kitaev limit are relevant for material realizations~\cite{Xu2018,PhysRevLett.123.037203,PhysRevResearch.3.013216}.  However, consensus is still lacking on whether the isotropic spin-1 model is gapless or gapped~\cite{JPSJ.87.063703,PhysRevB.102.121102,PhysRevResearch.2.022047,PhysRevB.105.L060403,Jahromi2024,PhysRevB.110.035121}, as well as on the existence of bond orders~\cite{PhysRevResearch.2.033318,PhysRevB.105.L060403,Jahromi2024,PhysRevB.110.035121} and topological order. For larger spin up to $S=2$, the results reported in Refs.~\cite{Jin2022,PhysRevB.108.075111,Jahromi2024} are not in agreement with the tripled-unit cell obtained in semiclassical results. Accurate numerics of both the spin-1 and the spin-$\tfrac{3}{2}$ model by exact diagonalization and density-matrix renormalization group report large finite-size effects; scaling up the size is extremely challenging.
Two-dimensional tensor network (TN) methods, specifically infinite projected entangled-pair states (iPEPS)~\cite{verstraete2004}, are not limited by such finite-size effects. Hence, iPEPS results for the higher-spin Kitaev models, obtained using improved optimization algorithms and more systematic data analysis that go beyond earlier studies based on imaginary-time evolution~\cite{PhysRevResearch.2.033318,PhysRevB.105.L060403,Jahromi2024}, are highly desirable.

\begin{figure}[tbp]
    \centering
    \includegraphics[width=0.45\textwidth]{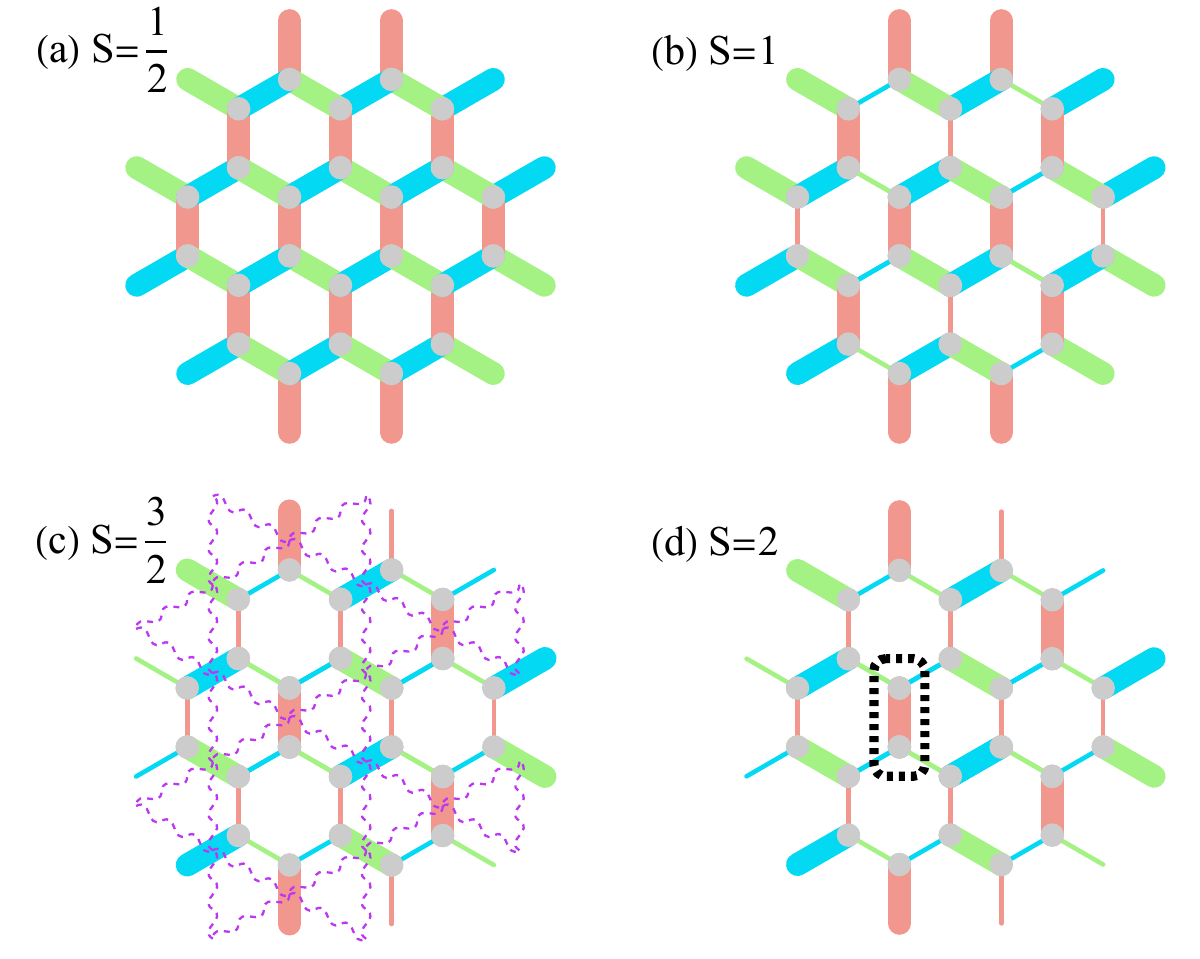}
    \caption{Ground-state bond-energy configurations in the Kitaev model for different spin values. Different colors represent spins along the $x$, $y$, and $z$ axes, and line thickness indicates the relative bond-energy strength. (a) Spin-$\tfrac{1}{2}$: no bond order, exhibiting uniform bond energies. The ground state is a gapless Kitaev spin liquid. (b) Spin-1: plaquette order. Each site is connected to one weak and two strong bonds. (c)~Spin-$\tfrac{3}{2}$: topological dimer order. Each site connects to one strong and two weak bonds. Purple dashed lines mark entanglement between strong bonds, forming a kagome lattice. (d)~Spin-2: trivial dimer order. Each site connects to one weak and two strong bonds, where the strong bonds form short-range entanglement. A  limiting case is a configuration where  one spin in  a bond (e.g., within the black dashed box) only entangles with the other spin: $\tfrac{1}{\sqrt{2}}\left(|\uparrow\uparrow\rangle+|\downarrow\downarrow\rangle\right)$, where the up and down arrows represent the largest and smallest $S_z$ components, respectively.}
    \label{fig: bond_configuration}
\end{figure}

In this Letter, we address the question of the existence and nature of valence-bond orders for Kitaev models with moderately higher spin $S \in (1, \tfrac{3}{2}, 2)$. We reveal that, for these higher-spin models, translational symmetry is broken with unit cells being tripled by forming valence-bond orders.  However, three distinct phases are realized for the different spin values, namely nontopological plaquette order (spin 1), topological dimer order (spin $\tfrac{3}{2}$), and nontopological dimer order (spin 2). These phases are illustrated in Figs.~\ref{fig: bond_configuration}(b)--\ref{fig: bond_configuration}(d). These findings are obtained with high-performance iPEPS simulations and complemented with an analytical argument showing that topological order is present for half-integer $S$ and absent for integer $S$ as a consequence of the properties of the conserved flux operators. 

In line with the ongoing development of variational iPEPS techniques~\cite{VPEPS,PhysRevB.94.155123,ADPEPS}, we employ gradient descent optimizations, integrating further algorithmic improvements to achieve significantly higher precision with iPEPS at large bond dimensions.   For the spin-$\tfrac{1}{2}$ Kitaev model~\cite{PhysRevB.90.195102,PhysRevLett.123.087203,PhysRevB.107.054424,PhysRevB.108.085103,QRC3v}, the transition from traditional imaginary-time evolution to modern gradient-descent schemes has pushed the ground-state energy precision to $\sim 10^{-7}$.  
Building on this progress, we use gradient-optimized iPEPS with enlarged unit cells to determine the translational symmetry breaking order of  higher-spin Kitaev models. Further evidence is acquired by contrasting these results with symmetry-restricted iPEPS results, where we propose a novel methodology to detect cat-state behavior associated with the translational symmetry breaking.
We furthermore characterize topological order through the PEPS transfer-matrix spectrum~\cite{haegeman2015shadows}. Unlike several previous works~\cite{SCHUCH20102153,haegeman2015shadows,PhysRevResearch.2.033318,PhysRevB.105.L060403}, we do not impose a $\mathbb{Z}_2$ virtual symmetry on the iPEPS but find that it emerges spontaneously~\cite{PhysRevB.101.041108,PhysRevB.101.115143} in the regime where topological order is expected. This is the case for spin-$\tfrac{3}{2}$. For the spin-1 model~\cite{PhysRevResearch.2.022047, PhysRevB.105.L060403}, our methodology avoids the problem of a cat state hiding the $\mathbb{Z}_2$ condensation. 

We start with a brief discussion of the Kitaev honeycomb Hamiltonian~\cite{KITAEV20062} generalized to arbitrary spin:
\begin{equation}
H = \sum_{\langle i,j \rangle_\gamma} K_\gamma S_i^\gamma S_j^\gamma ,
\end{equation}
where $\gamma$ denotes the $x$, $y$, $z$ bonds and $\left \langle i,j\right \rangle_{\gamma}$ denotes the nearest-neighbor pair on the $\gamma$ bond. The bonds of the honeycomb lattice are divided into $x$, $y$, $z$ sets, colored dark blue, yellow, and red in Fig.~\ref{fig: bond_configuration}. We focus on the isotropic case $K_x=K_y=K_z=K=-1$. The sign of $K$ can be flipped by four-unit-cell local $\pi$ rotations of spins~\cite{rousochatzakis2015phase}. Our conclusion will equally apply to the $K=1$ models because such rotations do not alter the physical quantities characterizing the quantum phases that we find. For each unit honeycomb plaquette, there is a locally conserved quantity $W_p=e^{i\pi(S^x_1+S^y_2+S^z_3+S^x_4+S^y_5+S^z_6)}$~\cite{KITAEV20062,PhysRevB.78.115116}. Because of Elitzur's well-known no-go theorem~\cite{PhysRevD.12.3978}, these local flux symmetries cannot be broken in the ground state. The global symmetry of $\pi$-spin rotations around the $x$, $y$, and $z$ axes can be composed from subsets of $W_p$ operators, thereby also precluding magnetic order in the ground state(s). 

\begin{figure}[tbp]
    \centering
    \includegraphics[width=0.44\textwidth]{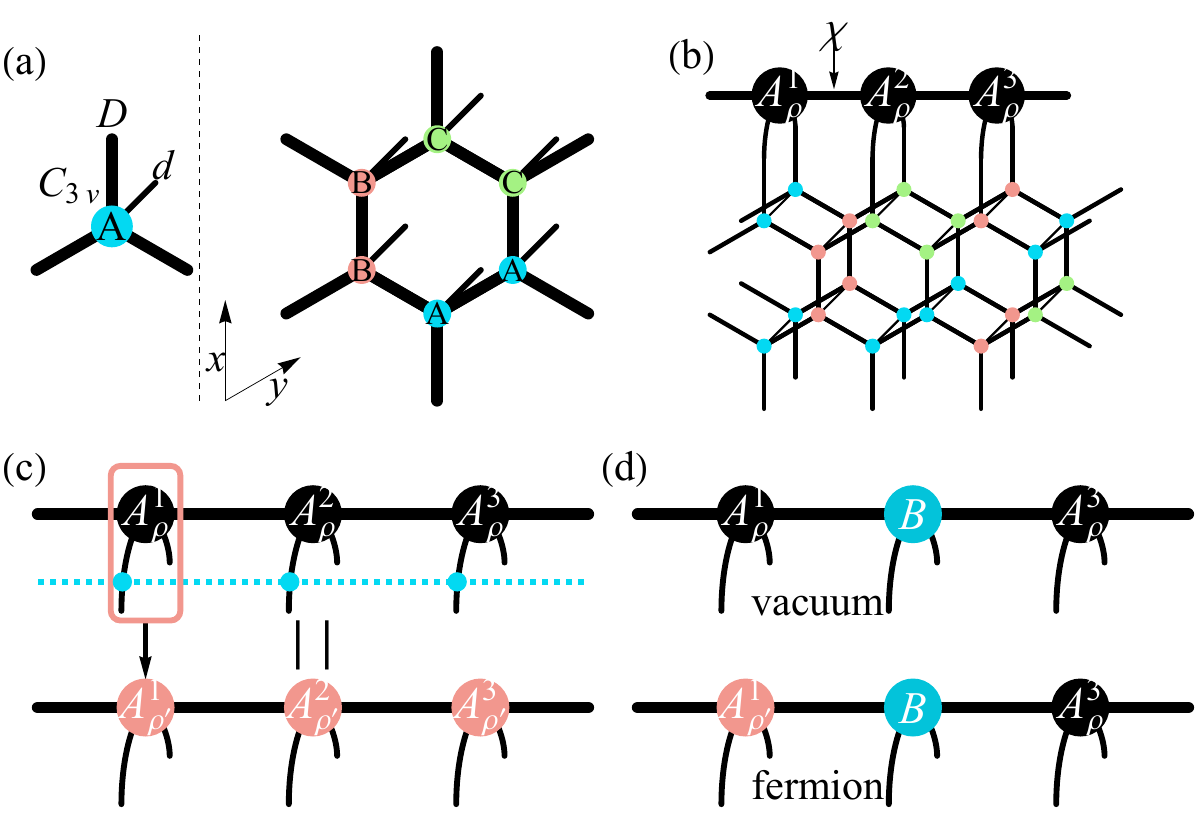}
    \caption{iPEPS ansatz and MPS boundary solutions in VUMPS contraction. (a) Left:  $C_{3v}$-symmetric tensor for single-tensor iPEPS. The tensor on the other sublattice is identical as a function of physical leg (dimension $d$) and virtual bonds $x$, $y$, and $z$ (dimension $D$). Right: tripled-unit-cell structure with three independent tensors (ABC). (b) Boundary MPS as the fixed-point solutions of the iPEPS transfer matrix (appearing in the norm), where the physical legs of the bra and ket have been contracted. The MPS is defined by the contraction of local tensors $A^{i}_\rho$ along the virtual legs with dimension $\chi$. (c) For $\mathbb{Z}_2$ topologically ordered states, the iPEPS transfer matrix exhibits a doubly degenerate fixed point structure, represented by boundary MPS with tensors $A_\rho$ and $A_{\rho^\prime}$ and associated with the breaking of an emerging virtual $\mathbb{Z}_2$ symmetry. (d) Transfer matrix spectrum then contains trivial excitations, targeted by the ansatz in the upper panel, as well as domain wall excitations between the two fixed points, targeted by the lower ansatz.}
    \label{fig: ipeps_unit_cell}
\end{figure}

To obtain the ground states of the spin-1, $\tfrac{3}{2}$, and 2 Kitaev  models, we perform gradient-descent optimization of iPEPS. The iPEPS are a family of variational wave functions in the thermodynamic limit, defined by a repeated unit cell of tensors, as depicted in Fig.~\ref{fig: ipeps_unit_cell}(a). The number of variational parameters scales as $D^3$, with $D$ being the bond dimension of the virtual bonds connecting the site tensors to their nearest neighbors [Fig.~\ref{fig: ipeps_unit_cell}(a)]. To directly identify translational symmetry breaking, we explore iPEPS with different enlarged unit cells and determine the tripled structure [Fig.~\ref{fig: ipeps_unit_cell}(b)] as the optimal choice. Physical observables of iPEPS such as bond energies are computed by controlled approximate contractions. We developed the variational uniform matrix product state (VUMPS)~\cite{PhysRevB.97.045145, 10.21468/SciPostPhysLectNotes.7, Nietner2020efficient, PhysRevB.108.085103} method to directly work with iPEPS on the honeycomb lattice with a general unit cell. The VUMPS method aims to solve the fixed-point environments of the  iPEPS as the top, bottom, up, and down boundary matrix product state (MPS). Parallel to the general unit cell calculations, we also use a symmetry-restricted iPEPS that is invariant under the full lattice $C_{3v}$ and translational symmetry---contracted using the $C_{3v}$ symmetric QR decomposition (QR) corner transfer matrix renormalization group (CTMRG) method~\cite{Nyckees2023,CTM,PhysRevB.107.054424,QR,QRC3v}---in order to provide a cross-check and to facilitate connecting to previous works. Both the VUMPS and QR-CTMRG methods are characterized by the environment bond dimension $\chi$, with increasing $\chi$ yielding higher precision. Energy gradients for the optimization are computed by automatic differentiation (AD)~\cite{ADPEPS,PhysRevB.108.085103,QRC3v} of VUMPS and QR-CTMRG, respectively. 

\begin{figure}[t]
    \centering
    \includegraphics[width=\linewidth]{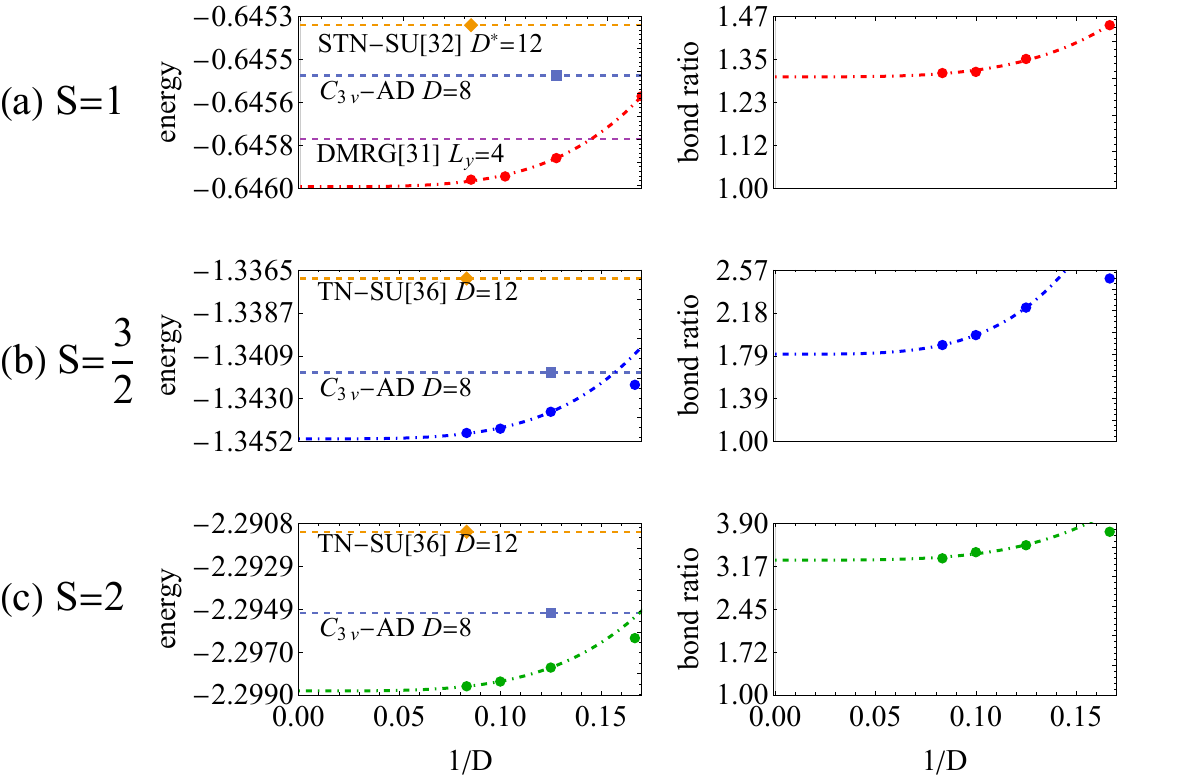}
    \caption{Average ground-state energy per site (left) and bond ratio (right) for higher-spin Kitaev models for (a) spin 1, (b) spin $\tfrac{3}{2}$, and (c) spin 2. Our results are plotted as points on extrapolated dashed lines ($1/D^\alpha$ extrapolation~\cite{suppm}) for unrestricted tripled-unit-cell iPEPS (see \cref{fig: ipeps_unit_cell}), with bond dimension $D$ up to 12, and as a square box for the $C_{3v}$-symmetric PEPS with $D=8$.    The contraction dimension $\chi$ for the environment is ramped to be large enough for convergence  (up to 512)~\cite{suppm}. Previous (symmetric) iPEPS results obtained using simple-update [(S)TN-SU]~\cite{PhysRevResearch.2.033318,Jahromi2024} as well as density matrix renormalization group (DMRG) results~\cite{PhysRevB.102.121102} are also indicated. The bond ratio is defined as the ratio between strong and weak bond energies $|\langle S^{\gamma}_i S^{\gamma}_j\rangle|_{\max}/|\langle S^{\gamma'}_{i'} S^{\gamma'}_{j'}\rangle|_{\min}$.
}
    \label{fig: config_ABC_results}
\end{figure}

We optimize iPEPS with different bond dimensions $D$ to approximate the ground state for the spin-1, $\tfrac{3}{2}$, and 2 models and summarize the results in Fig.~\ref{fig: config_ABC_results}. In the left panels, we plot the average energy per site obtained with both the unrestricted PEPS with tripled unit cell, as well as with the $C_{3v}$-symmetric and translational invariant PEPS. For comparison, we also include---to our knowledge---the best results for the energy per site currently available in the literature. We are able to consistently achieve lower variational energies with the symmetry-unrestricted iPEPS with a tripled unit cell. We verify that  $|W_p|=1$ up to $10^{-7}$ precision. 
Our results exhibit, to good approximation, only two distinct values of the bond energies. The bond ratio is thus defined as the magnitude of the stronger bond energy divided by that of the weaker one, as obtained from the tripled-unit-cell data, and is plotted in the right panels of Fig.~\ref{fig: config_ABC_results}.

The deviation of the bond ratio from 1 indicates the existence of valence-bond solid orders. The spin-$\tfrac{3}{2}$ and spin-2 states both exhibit the tripled-unit-cell dimer bond-energy configuration [Figs.~\ref{fig: bond_configuration}(c) and \ref{fig: bond_configuration}(d)] expected in the semiclassical limit, which features isolated strong bonds, i.e., dimers. We show below that only the spin-$\tfrac{3}{2}$ ground state exhibits topological order. For spin-$\tfrac{3}{2}$ and 2, Ref.~\cite{Jahromi2024} obtained anisotropic bond ordered states that do not break translational symmetry using iPEPS, optimized with imaginary-time evolution using simple update (TN-SU). Our results show that the tripled-unit-cell states are energetically more favored. For the spin-1 model, our results exhibit a different bond energy configuration resulting in plaquette order: connected strong bonds form a unit honeycomb plaquette on the lattice [Fig.~\ref{fig: bond_configuration}(b)]. In Supplemental Material~\cite{suppm}, we provide supporting evidence from finite-cluster results with suitable geometry. Our spin-1 energy density is lower than previous results~\cite{PhysRevB.102.121102,PhysRevResearch.2.022047}, including a DMRG calculation for an infinite cylinder geometry with a 4-unit vector width~\cite{PhysRevB.102.121102}
and an iPEPS calculation using imaginary time evolution with a simple update that preserves the translational, $C_{3v}$, and flux conservation symmetries (STN-SU)~\cite{PhysRevResearch.2.033318}. The latter result only differs by $\sim 0.001$ with our gradient-optimized symmetry-restricted calculation ($C_{3v}$-AD), thus raising the question of whether both results are capturing the same phase, which may furthermore be compatible with the symmetry breaking that is explicitly present in our symmetry-unrestricted results. 

\begin{figure}[tbp]
    \centering
    \includegraphics[width=0.48\textwidth]{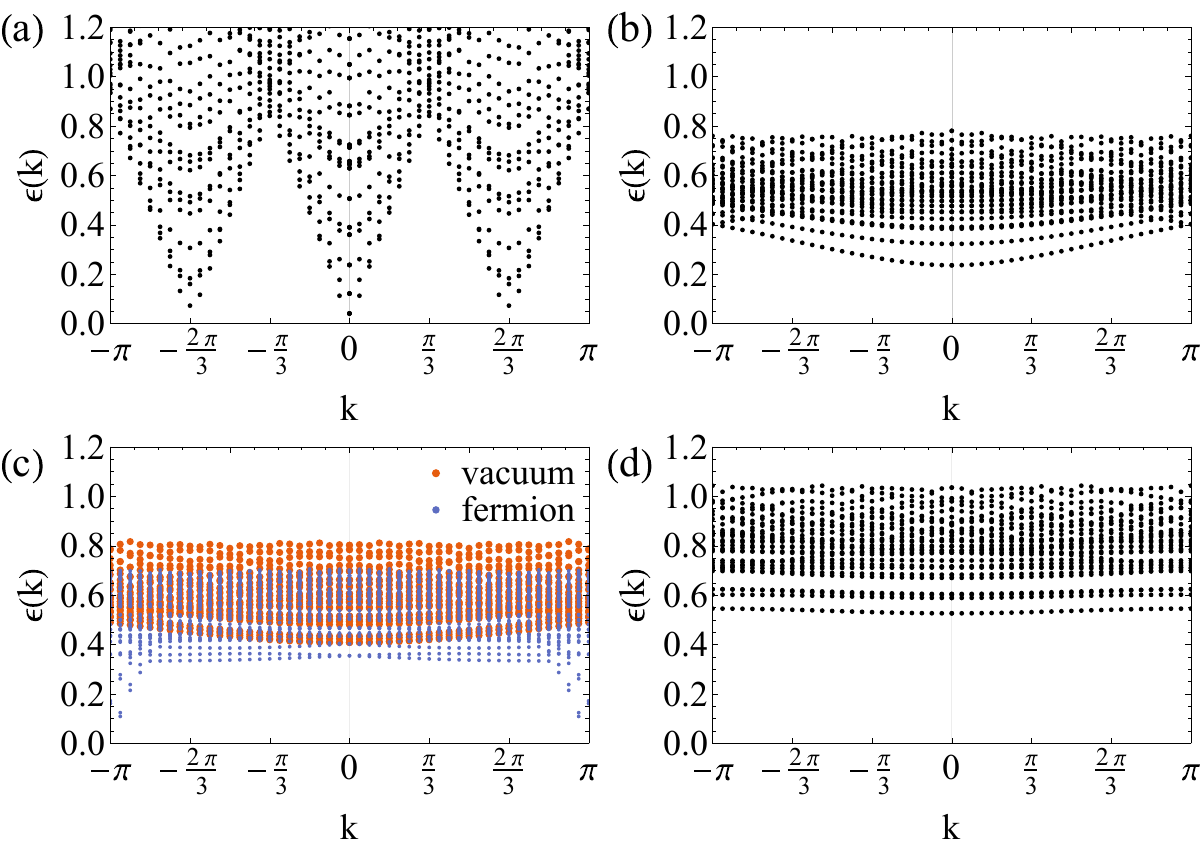}
    \caption{Spectra of inverse correlation lengths [see Eq.~\eqref{eq:inversecorr}] obtained from the transfer matrix for the single-tensor unit cell  $C_{3v}$-restricted iPEPS for the spin-$1$ model (a), the tripled-unit-cell iPEPS for spin-1 (b), spin-$\tfrac{3}{2}$ (c), and spin-2 (d). The transverse momentum along the transverse matrix corresponds to a cut in the Brillouin zone (a) or in the threefold reduced Brillouin zone (b)--(d), where $(\pm 2\pi/3)$ corresponds to (reduced) K and K' points. Only for spin-$\tfrac{3}{2}$ (c) do we find two degenerate boundary MPS, for which we can plot both a vacuum and a fermion spectrum, using the ansatz in Fig.~\ref{fig: ipeps_unit_cell}(d).}
    \label{fig: TM_spectrum}
\end{figure}

Indeed, while the symmetry-restricted iPEPS results (STN-SU \cite{PhysRevResearch.2.022047} and our own $C_{3v}$-AD results) cannot explicitly break the translational symmetry due to the single-tensor unit cell, it is possible that they encode a cat-state superposition of bond-ordered states.
Following standard practices~\cite{PhysRevLett.61.365}, we determine whether the bond-bond correlation function exhibits long-range order. As shown in  Fig.~\ref{fig:dimer_correlation_FCLS}, 
the data indicates that the tripled-unit-cell oscillation may survive in the long-range limit. The behavior is different from that of the spin-$\tfrac{1}{2}$ case, where an algebraic decay with exponent 4 has been determined~\cite{PhysRevA.78.012304,PhysRevResearch.2.013005,QRC3v}. Detecting cat-state superpositions from correlation functions is standard practice in many other numerical methods. However, using the iPEPS transfer matrix, we can probe this behavior even more explicitly. The momentum-resolved transfer-matrix spectra, parameterized by the transverse momentum $k_\parallel$ as
\begin{equation}
\lambda_n(k_\parallel) = \exp\left[-\epsilon_n(k_\parallel) + \mathrm{i} \kappa_n(k_\parallel)\right],\label{eq:inversecorr}
\end{equation}
directly encodes both the decay rate and the oscillatory behavior of static two-point correlation functions~\cite{haegeman2015shadows,Zauner_2015,PhysRevX.8.041033}, where the angular frequency $\kappa(k)$ encodes the longitudinal momentum $k_\perp$ and $\epsilon(k)$ can be interpreted as an inverse correlation length and serves as a proxy for the physical excitation energies in the system. We plot the spectra $\epsilon(k)$ of the symmetry-restricted iPEPS transfer matrix for the spin-1 model in Fig.~\ref{fig: TM_spectrum}(a). A gap closing tendency at $K$ and $K'$ indicates either a slowly decaying oscillation or long-range order~\cite{PhysRevB.92.155133}, compatible with the symmetry broken state obtained in the tripled unit cell. To relate this gap closing to a specific order, we use an enlarged unit cell environment to contract the $C_{3v}$-AD iPEPS, after it has been optimized in a translation invariant environment.
 We then find that the plaquette and dimer orders appear (Fig.~\ref{fig: bond_configuration}), 
consistent with our enlarged unit cell iPEPS results. This analysis thus provides more detailed information than the bond-bond correlation function that we calculated to distinguish these two orders. Further details are provided in the End Matter.

The transfer matrix also encodes the presence or absence of topological orders. Hereto, we first determine the number of dominant eigenvalues, i.e., distinct VUMPS boundary MPS solutions. Starting from random initializations, we find the twofold degeneracy characteristic of $\mathbb{Z}_2$ topological order for the  spin-$\tfrac{3}{2}$ model, whereas unique boundary states are obtained for the spin-$1$ and spin-$2$ cases. This is consistent with our theoretical analysis, discussed later, which explains the odd-even $2S$ discrepancy for the presence of topological order. The occurrence of different boundary states can be directly related to the existence of 1-form (i.e., stringlike) symmetries in topologically ordered states.
Absorbing a virtual string in the boundary environment must give rise to another degenerate boundary solution, in order to support deconfined anyonic excitations \cite{haegeman2015shadows}, as depicted in Figs.~\ref{fig: ipeps_unit_cell}(c) and \ref{fig: ipeps_unit_cell}(d). 

With the boundary solutions at hand, we now also plot the transfer matrix spectra of $\epsilon(k)$ for unrestricted iPEPS with the tripled unit cell in Figs.~\ref{fig: TM_spectrum}(b)--\ref{fig: TM_spectrum}(d). Gapped spectra for $\epsilon(k)$ are observed for the spin-$1$, $\tfrac{3}{2}$, and $2$ models, indicating the existence of physical excitation gaps.  As we find two degenerate boundary MPS for the spin-$\tfrac{3}{2}$ case, a domain wall structure can be used to construct a transfer matrix that encodes the string correlation [see \cref{fig: ipeps_unit_cell}(d)], which corresponds to the two-point correlation of emergent fermions.  

Now we provide a theoretical analysis to explain the absence and presence of topological order in the case of integer and half-integer spin, respectively. The idea is to examine the flux operators.  For dimer orders, one can construct a prototype of short-range-entangled states by maximally entangling the spins in a dimer: $\tfrac{1}{\sqrt{2}}\left(|\uparrow\uparrow\rangle+|\downarrow\downarrow\rangle\right)$ for $K=-1$ and $\tfrac{1}{\sqrt{2}}\left(|\uparrow\downarrow\rangle+|\downarrow\uparrow\rangle\right)$ for $K=1$. A straightforward calculation shows that a product state of such dimers is an eigenstate of the flux operators for integer spins but not for half-integer spins. This applies to all dimer coverings.  The nontopological decoupled plaquette state can also be constructed. For plaquette orders, the decoupled plaquette limit can be constructed by turning off the couplings at the weak bonds in the pattern. We solve an isolated plaquette and find that the ground state is unique for spin 1~\cite{suppm}.  The above constructions show that, for spin 1 and spin 2, there is a nontopological state with the same broken symmetries, indicating a possible adiabatic connection between the optimized iPEPS and these prototypical zero-correlation-length states. 

Furthermore, we consider an effective theory for strong tripled-unit-cell dimer orders. As mentioned before, the Kitaev type of spin-spin interaction per bond is Ising-like (e.g., $K_x S_i^x S_j^x$) with two degenerate spin orientations. An effective theory can be constructed by assigning a spin variable $\sigma^z$ with a basis of two low-energy states of the bond. The construction gives an effective spin-$\tfrac{1}{2}$ kagome lattice [Fig.~\ref{fig: bond_configuration}(c)] model~\cite{Rousochatzakis2018}. To reveal the discrepancy between the two classes, we find that it is enough to consider the restriction put to possible effective terms from the flux conservation. For half-integer spin cases, one can construct a Hamiltonian with $W_p$ symmetry:
\begin{align}\label{eq:kagometoric1}
H^{\mathrm{top}}_{\mathrm{eff}}= \sum_{i,j...n \in \hexagon} J_{e}\sigma^{x}_{i} \sigma^{x}_{j} \sigma^{x}_{k} \sigma^{x}_{l} \sigma^{x}_{m} \sigma^{x}_{n}  
+\sum_{a,b,c\in \triangle} J_{m}\sigma^{z}_{a} \sigma^{z}_{b} \sigma^{z}_{c},
\end{align}
where $i$-$n$ and $a$-$c$ label the sites of honeycomb and triangular plaquettes in the kagome lattice, respectively. The first and second terms are effective $W_p$ operators commuting with each other.  
This is the toric code model on the kagome lattice realizing $\mathbb{Z}_2$ topological order, which has been obtained before from semiclassical analysis~\cite{Rousochatzakis2018}.
For the integer $S$ case, we instead obtain a Hamiltonian:
\begin{align}\label{eq:kagometransverse}
H^{\mathrm{tri}}_{\mathrm{eff}}= \sum_{i} h \sigma^x_{i} ,
\end{align}
where $i$ labels the lattice sites. The Hamiltonian equation~\eqref{eq:kagometransverse} commutes with the effective $W_p$ operators $W_{p,i}=I$ on honeycomb and $W_{p,j}=\sigma^{x}_{j,a} \sigma^{x}_{j,b} \sigma^{x}_{j,c}$ triangle plaquettes. This stabilizes the trivial state of isolated dimers. 

The mechanism for valence-bond order formation beyond the semiclassical picture deserves discussion. Previous theoretical investigations of spin-$\tfrac{3}{2}$ Kitaev models have developed an interacting pseudofermionic theory~\cite{Jin2022, PhysRevB.108.075111} for spin-$\tfrac{3}{2}$. It has been found at the mean-field level, by assuming translational symmetry, that fermions form nearly flat bands with Majorana cones. We find that the spin bond order is translated into interaction-induced pseudofermion bond order, while the $\mathbb{Z}_2$ gauge structure is intact. For the spin-1 Kitaev model, as there is no obvious emergent fermion and braiding~\cite{PhysRevB.105.L060403,PhysRevResearch.2.022047}, previous works proposed candidate symmetric phases with a picture of anyonless $\mathbb{Z}_2$ spin liquids~\cite{PhysRevB.78.115116,PhysRevB.105.L060403,PhysRevResearch.2.022047,PhysRevLett.130.156701,10.21468/SciPostPhys.16.4.100}. Our Letter shows that the spin-1 Kitaev model forms trivial plaquette order. Since the order is moderate, certain 
additional interactions may help suppress the plaquette order to stabilize a possible nearby symmetric state. Adding sufficient explicit coupling anisotropy has been shown to result in a trivial columnar VBS state~\cite{JPSJ.87.063703,PhysRevB.99.104408,minakawa2020magnetic,PhysRevB.105.L060405}.

In conclusion, we have demonstrated that the higher-spin Kitaev models provide a platform for an intriguing interplay among valence bond orders and spin-liquid concepts. Considering additional terms relevant for Kitaev materials can lead to rich phenomena, such as interplay between bond and magnetic order and between anisotropy and translational symmetry breaking. Furthermore, our Letter provides an improved tensor network workflow for numerical investigations of symmetry breaking and topological order in frustrated magnets. 

\paragraph{Acknowledgments---}
We thank Nick Bultinck, Naichao Hu, Rui-Zhen Huang, Shenghan Jiang, Johannes Knolle, Sid Parameswaran, Illya Lukin, Hong-Hao Tu, Frank Verstraete, Victor Vanthilt, Kang Wang, Wei Tang, Fan Yang, Xu Zhang, and Zheng Zhu for discussion. X.Y.Z., J.H., and Y.H. are supported by the European Research Council under the European Union's Horizon 2020 program under Grant Agreement No.~101125822. Q.Y. and P.C. are supported by the European Research Council under the European Union's Horizon 2020 program under Grant Agreement No.~101001604. For code development and numerical calculations, we are thankful for the use of the University of Oxford Advanced Research Computing facility, and we thank the EuroHPC Joint Undertaking for access to MareNostrum5 at BSC, Spain, under Grants No.~EHPC DEV-2025D07-003 and No.~DEV-2025D06-052 and access to LUMI, Finland, under Grant No.~VSC-2025-09-T350-JH-JULIA.

Views and opinions expressed are, however, those of the authors only and do not necessarily reflect those of the European Union or the European Research Council Executive Agency. Neither the European Union nor the granting authority can be held responsible for them.

\paragraph{Data availability---}
The data that support the findings of this article are openly available~\cite{data}, embargo periods may apply.



\section*{End Matter}

\appendix
\setcounter{secnumdepth}{1}

\section{VUMPS algorithms for native honeycomb lattice}\label{VUMPS}
To compute gradients and physical observables, we employ two contraction schemes depending on the iPEPS unit cell. For the translationally invariant and $C_{3v}$-symmetric iPEPS, we use the efficient QR-CTMRG method~\cite{QRC3v}, while for enlarged unit cells we adopt the VUMPS algorithm. Both approaches offer a computational advantage over the original CTMRG in obtaining fixed-point environments. Although randomized singular value decomposition~\cite{PhysRevB.107.054424} can reduce the cost of CTMRG to a similar level, QR-CTMRG and VUMPS remain faster in practice. Our QR-CTMRG implementation is adapted to the honeycomb lattice and exploits its $C_{3v}$ symmetry for further speedup. As QR-CTMRG is not yet generalized to enlarged unit cells, VUMPS is used in those cases, as well as for cross-validation and analysis of $C_{3v}$-symmetric states with larger environment cells, as discussed in the main text. Details of the VUMPS can be found in Ref.~\cite{PhysRevB.108.085103}, and our additional general performance improvements will be reported elsewhere. For gradient optimization, we implement a precondition scheme~\cite{Zhang2026precondition}.

\begin{figure}[tbp]
    \centering
    \includegraphics[width=0.25\textwidth]{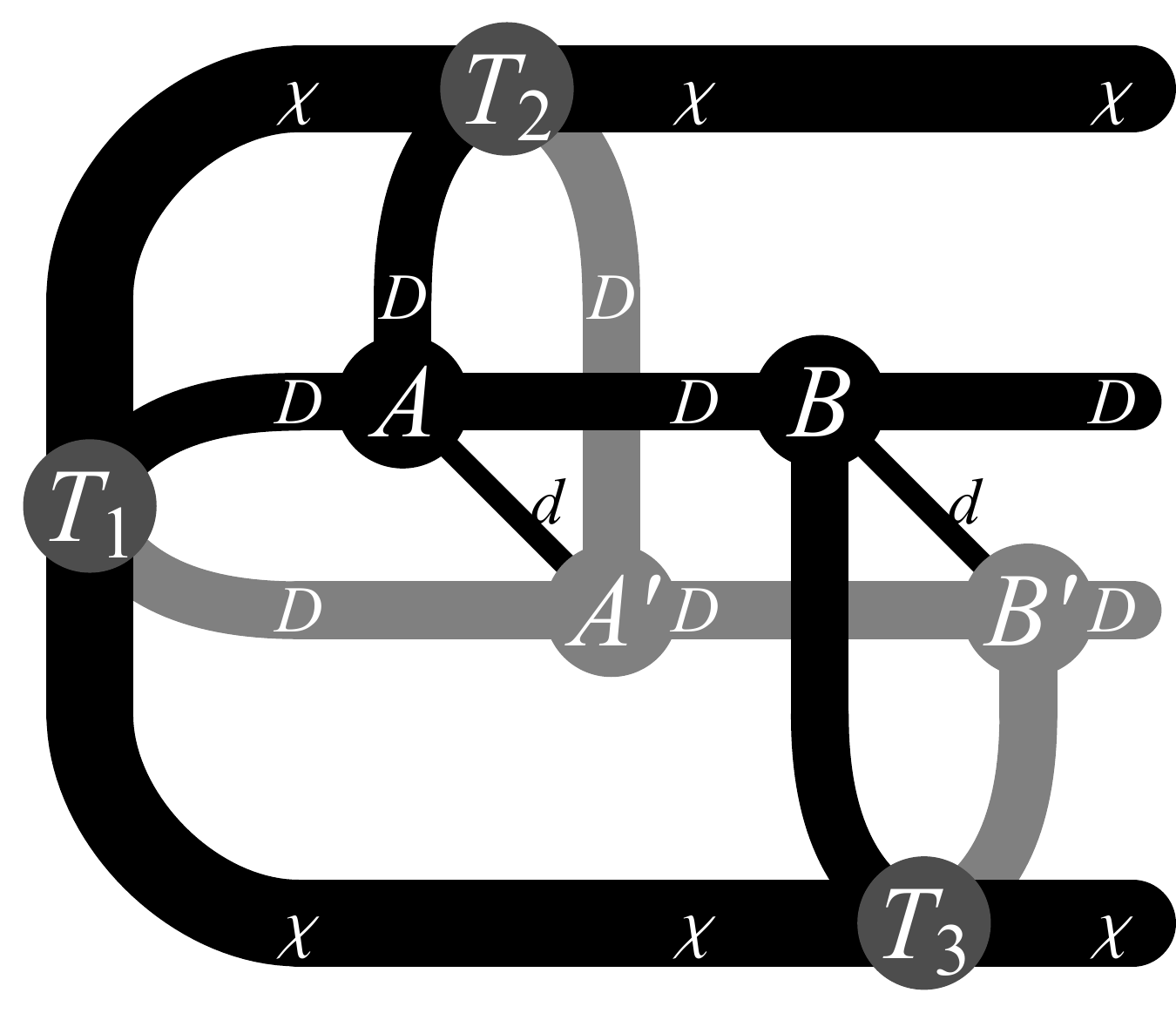}
    \caption{Linear transfer matrix map: the optimal contraction order is $T_1 \rightarrow T_2 \rightarrow A \rightarrow A^\prime \rightarrow T_3  \rightarrow B  \rightarrow B^\prime$, and the corresponding computational costs are $O(\chi^3D^4) \rightarrow O(\chi^2D^5d) \rightarrow O(\chi^2D^5d) \rightarrow O(\chi^3D^4) \rightarrow O(\chi^2D^5d) \rightarrow O(\chi^2D^5d)$.}
    \label{fig: TM_contraction}
\end{figure}

We focus on the efficient contraction of native honeycomb iPEPS using the VUMPS algorithm. In contraction, the bra-ket network surrounding a lattice site is approximated by environment tensors (denoted as $ T $ in \cref{fig: TM_contraction}). The primary computational cost in VUMPS arises from applying the linear transfer matrix map. The depicted contraction scheme scales as $ O(\chi^3D^4 + \chi^2D^5d) $; $ C_{3v} $ CTMRG exhibits the same scaling when using a similar method. A key advantage of contracting the honeycomb lattice natively is computational efficiency. A common alternative is to merge two sites to form a square lattice, but this is effectively a more costly contraction order. The merged square iPEPS tensor (with four virtual bonds) has a higher contraction cost of $ O(\chi^3D^4 + \chi^2D^6d^2) $. Assuming the typical empirical relation $ \chi = O(D^2) $, both methods ultimately scale as $ O(D^{10}) $. However, working directly with the honeycomb lattice reduces the prefactor, offering a practical speedup.

\section{Inferring cat states from symmetry-restricted iPEPS}\label{catinference}
In the main text, we demonstrate that signatures of symmetry breaking can be inferred from symmetry-restricted iPEPS via cat-state behavior. Here we provide further details and supporting data.

\begin{figure}[tbp]
    \centering
    \includegraphics[width=0.45\textwidth]{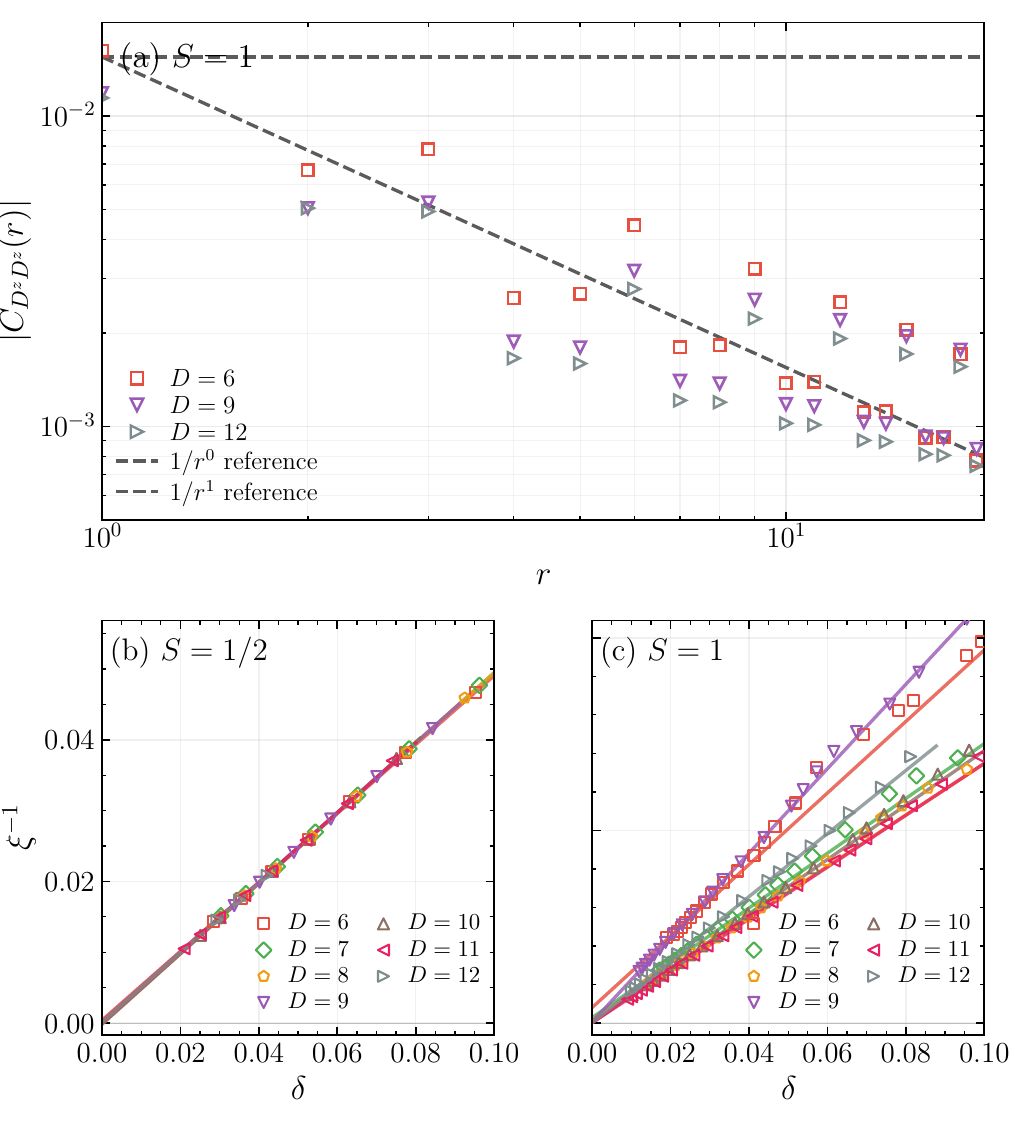}
    \caption{(a) Bond-bond correlation function of symmetry-restricted iPEPS for $S=1$. The horizontal dashed line depicts the plateau behavior expected for oscillatory long-range orders, and the other reference dashed line depicts the form of $1/r$ algebraic decay.  More accurate data (larger $D$) show that the tripled-unit-cell oscillation may converge to plateaus.  (b),(c) Finite correlation length scaling for $S=\tfrac{1}{2}$ (b) and $S=1$ (c), where $\delta=\ln(|\lambda_1/\lambda_3|)$ with $\lambda_i$ being the $i$th largest-magnitude eigenvalue of transfer matrices and $\xi^{-1}=-\ln(|\lambda_1/\lambda_0|)$ is the inverse dominant correlation length.}
    \label{fig:dimer_correlation_FCLS}
\end{figure}

\paragraph{Bond-bond correlation functions:} In the thermodynamic limit, a cat state associated with the expected tripled-unit-cell symmetry-breaking pattern corresponds to a superposition of three degenerate states, yielding long-range order in specific connected correlation functions. For valence-bond order, the relevant quantity is the bond-bond correlation function,
\begin{align}\label{eq:bondbond}
C_{D_z,D_z}(r_{i,j}) = \langle S^z_i S^z_{i'} S^z_j S^z_{j'} \rangle
- \langle S^z_i S^z_{i'} \rangle \langle S^z_j S^z_{j'} \rangle,
\end{align}
where $i,i'$ and $j,j'$ denote nearest-neighbor pairs and $r_{i,j}$ is the distance between sites $i$ and $j$.
Based on our optimized $C_{3v}$ iPEPS, correlations can be reliably inferred over finite ranges, which extend as $D$ increases. Since long-range order is defined only in the infinite-distance limit, its emergence must be inferred from the trend with increasing $D$. Both tripled-unit-cell dimer and plaquette orders in the bond-bond correlation function manifest as oscillations with period three, taking values $(2c,-c,-c)$, where $c$ is a constant.  In variational iPEPS, correlation functions are generally more challenging to compute accurately than ground-state energies, with precision estimated from results at different bond dimensions $D$. Figure~\ref{fig:dimer_correlation_FCLS}(a) presents our spin-1 results. For $D=12$, the energy (nearest-neighbor correlation) error is about $10^{-4}$. The enlarged-unit-cell iPEPS yields a correlation length of roughly three lattice constants for the symmetry-broken state, suggesting that correlation data up to ten sites, accurate to about $10^{-3}$, can reliably indicate the ordering tendency. The expected oscillation is clearly observed, although the plateau shows a slight decay, which we attribute to numerical artifacts.
Similar finite-accuracy effects have been reported in large-scale PEPS studies of other spin models~\cite{LIU20221034}.  We observe that the estimated exponent tends toward zero with increasing $D$. By contrast, algebraic QSL is unlikely to have a very small exponent to remain stable~\cite{LIU20221034}.  (For spin-$\tfrac{1}{2}$, the exponent is 4.) If interpreted as algebraic decay, the extracted exponent is significantly smaller than 1, rendering the QSL scenario unlikely.

\paragraph{Transfer matrix spectrum:}
The long-range order in correlation functions is formally associated with the exact degeneracy of the dominant eigenvalues in the boundary MPS transfer-matrix spectrum~\cite{PhysRevB.92.155133}. In practice, iPEPS optimized at finite bond dimension $D$ need not exhibit exact degeneracy, but our results show a clear convergence toward it with increasing $D$. This behavior corresponds to the suppression of the apparent artificial decay in correlation functions. A static connected two-point correlation function, including Eq.~\eqref{eq:bondbond}, can be expressed as
\begin{align}
f(r) = \sum_{i>0} f_i e^{-\epsilon_i r}, \quad \epsilon_i = -\ln(\lambda_i / \lambda_0),
\end{align}
where $\lambda_i$ are the eigenvalues of the transfer matrix determined by the MPS bond dimension $\chi$ for a given iPEPS bond dimension $D$. Accurate estimation of the spectrum requires a simultaneous increase of both $\chi$ and $D$. The approach of the inverse dominant correlation length $\xi^{-1} = \ln(|\lambda_1 / \lambda_0|) \to 0$ signals that the connected correlation function ceases to decay exponentially.
As shown in Figs.~\ref{fig:dimer_correlation_FCLS}(b) and \ref{fig:dimer_correlation_FCLS}(c), we perform finite-correlation-length scaling following Refs.~\cite{PhysRevX.8.041033,QRC3v}. For spin~1, the extrapolated $\xi$ ($\chi \!\to\! \infty$, $\delta \!\to\! 0$) appears to converge for fixed $D$ yet remains relatively large even at moderate $D$. This behavior is consistent with Refs.~\cite{PhysRevResearch.2.033318,PhysRevB.105.L060403}, which achieve comparable energy accuracy to our $D=5$, $6$ data. We further find that $\xi$ tends to increase with $D$, suggesting an approach toward degeneracy consistent with cat-state behavior. To contrast this with a known algebraic QSL, we compare to the $S=\tfrac{1}{2}$ case: for $S=\tfrac{1}{2}$, data with $D \ge 6$ collapse, whereas for $S=1$, no such collapse occurs up to $D=12$. Moreover, the slope decreases with increasing $D$ for both even and odd $D$, indicating that, at fixed $\delta$, the gap between the leading and first excited transfer-matrix eigenvalues closes more rapidly as $D$ grows.

\begin{figure}[tbp]
    \centering
    \includegraphics[width=0.45\textwidth]{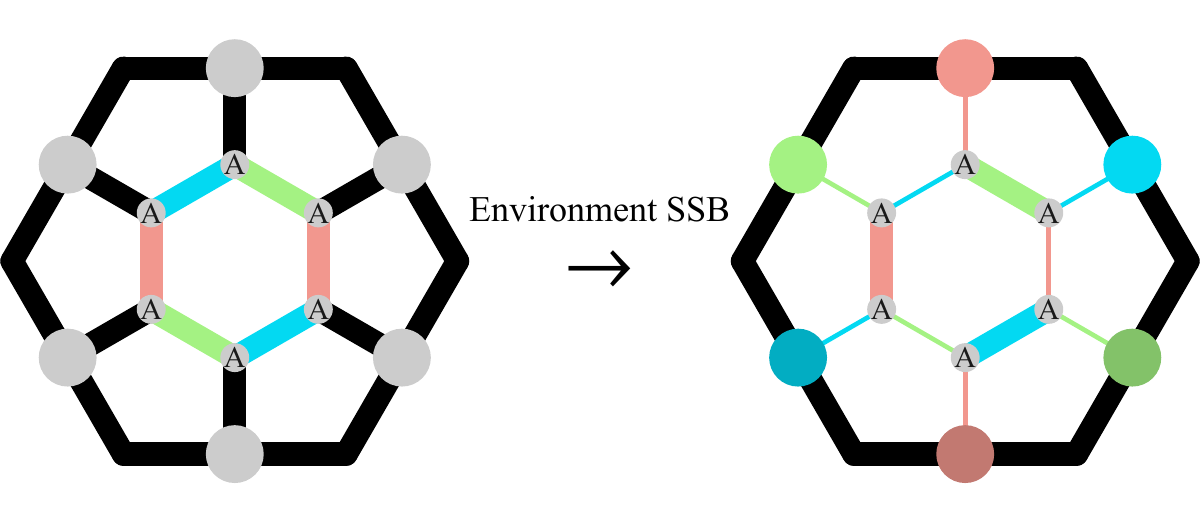}
    \caption{Spontaneous symmetry breaking of environment for contracting uniform $C_{3v}$ iPEPS tensor once the uniform restriction of environment is lifted. }
    \label{fig:enlarge_grid_SSB}
\end{figure}

\paragraph{Spontaneous symmetry breaking by environments:}
As a third approach, contracting the optimized symmetry-restricted iPEPS with an unrestricted environment directly demonstrates symmetry breaking. Explicitly enlarging the unit cells of the environments to $2\times6$ in combination with replicated PEPS tensors from the single unit cell, the calculated bond energies become nonuniform due to environmental symmetry breaking, as shown in \cref{fig:enlarge_grid_SSB}. This is a direct consequence of strong oscillatory bond-bond correlations. This method can directly reveal plaquette order and dimer order. When the degeneracy of transfer matrix dominant eigenvalues is not exact, the order exists only for finite $\chi$.


\section{Determine topological order from boundary MPS}\label{TO}
As shown in the main text, the action of string operators relating topological sectors leads to a twofold degeneracy in the boundary MPS of the iPEPS. Our approach employs unstructured site tensors and searches for distinct boundary MPS of the optimized iPEPS~\cite{PhysRevB.101.041108}. To explore different solutions, we use random initializations in the approximate boundary solver. Different boundary solutions produce energy densities that agree up to relative precision $10^{-9}$. Different solutions are identified by computing the inner-product density between six consecutive MPS tensors, with overlaps within $10^{-6}$ considered identical.
Unlike most previous works that impose $\mathbb{Z}_2$ virtual symmetry to realize topological degeneracy through virtual symmetry breaking, our method avoids such constraints. This is important because, in such symmetric constructions, even trivial phases can exhibit artificial fourfold boundary degeneracy---interpreted as condensations that break all $\mathbb{Z}_2 \!\times\! \mathbb{Z}_2$ symmetries of the iPEPS bra-ket structure~\cite{haegeman2015shadows}---which doubles the required bond dimension $D$ and increases the computational cost. The relationship between degeneracy and symmetry breaking becomes subtle when condensation coincides with translational symmetry breaking under restricted translational invariance. Following Ref.~\cite{PhysRevB.105.L060403}, we apply projectors~\cite{PhysRevLett.123.087203,PhysRevResearch.2.033318} to construct $\mathbb{Z}_2$-iPEPS from our $C_{3v}$ data. Random initialization then reveals that the degeneracy is more than twofold, with overlaps around $0.4$, indicating cat states of virtual symmetry. This is consistent with the absence of topological order in the $S=1$ phase.

\startSupplement
\hypersetup{citecolor=black}
\makeatletter
\@booleanfalse\force@deferlist@sw
\@booleanfalse\raggedcolumn@sw
\flushbottom
\begingroup
  \let\@AAC@list\arxiv@AAC@list
  \let\@AFF@list\arxiv@AFF@list
  \let\@AFG@list\arxiv@AFG@list
  \def\@title{Supplemental Material for ``Topological and Trivial Valence-Bond Orders in Higher-Spin Kitaev Models''}%
  \let\@date\@empty
  \let\mini@notes\@empty
  \setbox\absbox\box\voidb@x
  \let\label\@gobble
  \title@column\titleblock@produce
  \suppressfloats[t]
\endgroup
\makeatother

This Supplemental Material provides further details of the iPEPS simulations (\cref{sec:ipepsdetails,sec:chiramp,sec:extrapolation,sec:FCLS}), the exact solution of an isolated plaquette (\cref{sec:plaquette}), and additional exact diagonalization (ED) and density matrix renormalization group (DMRG) calculations supporting the plaquette order of the spin-1 model (\cref{sec:finitesize}). Figures and equations labeled with the prefix ``S'' refer to this Supplemental Material; all other figure numbers refer to the main text (including its End Matter), and reference numbers refer to the reference list of the main text.

\section{More details of iPEPS and contraction ansatz}\label{sec:ipepsdetails}
Our calculations and all of the referenced previous iPEPS work~\cite{PhysRevResearch.2.022047,PhysRevB.105.L060403,Jahromi2024} of higher-spin Kitaev models are based on honeycomb iPEPS, where the tensor on each site has three virtual legs, with the dimension set to be $D$. As the network is native to the lattice geometry, it brings no artifact for studying the lattice symmetry breaking of our interest. To investigate translational symmetry breaking beyond the scope of Refs.~\cite{PhysRevResearch.2.022047,PhysRevB.105.L060403,Jahromi2024}, we extend the translational symmetry of iPEPS from the minimal two-site unit to enlarged cases.

\begin{figure}[!ht]
    \centering
    \includegraphics[width=0.45\textwidth]{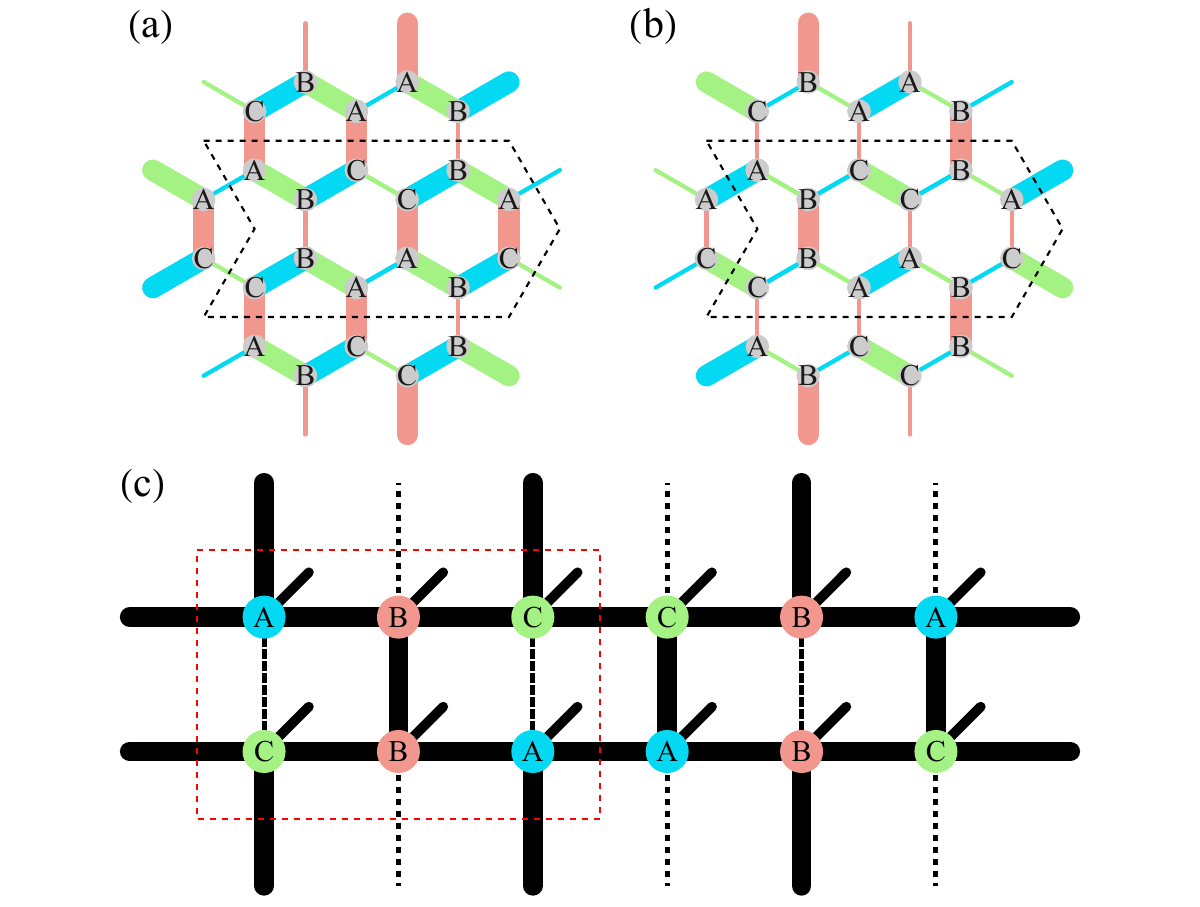}
    \caption{iPEPS ansatz with large unit cell. (a) Bond pattern for spin-1, (b) bond pattern for spin-$\tfrac{3}{2}$ and spin-2, (c) $2\times6$ unit cell with a brick wall shape. The dashed black lines indicate bonds omitted relative to the square-lattice embedding, and the dashed red box indicates the unit cell for the boundary MPS calculation.}
    \label{fig: ipeps_large_unit_cell}
\end{figure}

This process begins with the construction of a unit cell. The configuration presented in Fig.~\ref{fig: bond_configuration} of the main text has a unit cell with 6 sites. For VUMPS contraction, it is better to illustrate the assignment of the environment tensors in a further enlarged $2\times6$ unit cell (see \cref{fig: ipeps_large_unit_cell}). We draw the network alternatively by embedding the honeycomb lattice in a square lattice. The honeycomb iPEPS network remains unchanged because it is defined by the unchanged connection topology. The variational parameters are contained in the independent tensors $A$, $B$, and $C$, optimized by gradient descent. We set site tensors expected with the same bond energy $x$, $y$, $z$ distribution to be the same, a generalization of the single-tensor-for-two-sublattice restriction in Refs.~\cite{PhysRevLett.123.087203,QRC3v}. We verify the validity by lifting the restriction. However, we cannot fully reduce the number of environment tensors in the same spirit, as a converging contraction could not be achieved with the restrictions that we tried to impose. We resort to an environment composed of six boundary MPS tensors, surrounding the six tensors enclosed in the red dashed rectangle in \cref{fig: ipeps_large_unit_cell}(c), respectively. The environments for the parts not enclosed are obtained from translation.

\section{VUMPS \texorpdfstring{$\chi$}{chi}-ramp}\label{sec:chiramp}
\begin{figure}[!ht]
    \centering
    \includegraphics[width=0.5\textwidth]{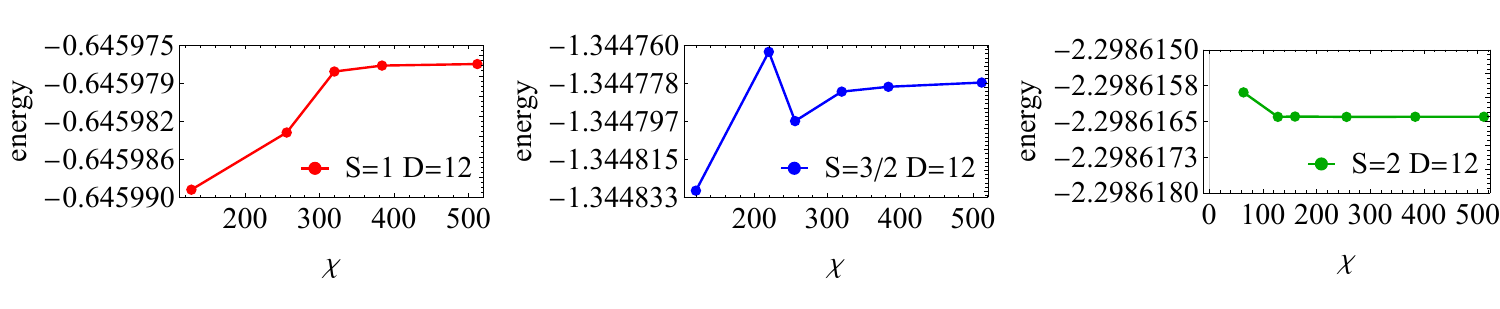}
    \caption{We ramp the environment bond dimension $\chi$ in the VUMPS algorithm to determine the energy of our optimized symmetry-breaking iPEPS. Here we show the data for $D=12$; the $\chi$ is ramped up to 512. The convergence hints that we achieve sufficient precision up to $10^{-6}$ or even $10^{-7}$.}
    \label{fig:chi_ramp_D12}
\end{figure}

High-precision evaluation of iPEPS observables requires a sufficiently large environment bond dimension; we have performed the ramp for the data that we presented in the main text. In \cref{fig:chi_ramp_D12}, we show an example of such a ramp.

\section{Extrapolation}\label{sec:extrapolation}
We perform a $1/D^\alpha$ extrapolation for both the symmetry-breaking iPEPS (large unit cell) and the symmetry-restricted iPEPS results. In Fig.~\ref{fig: config_ABC_results} of the main text, we present the extrapolated energy and bond ratio for the symmetry-breaking iPEPS. For $S=1$, the full dataset $D=6,8,10,12$ is used in the extrapolation, whereas for $S=\tfrac{3}{2}$ and $S=2$, only the data points with $D=8,10,12$ are included.

This choice is guided by the principle that a reliable extrapolation should be based on a well-converged sequence in $D$. First, the data for consecutive bond dimensions should vary smoothly rather than exhibit abrupt changes. As an illustration, we examine the correlation spectrum (Fig.~\ref{fig: TM_spectrum} of the main text) for the spin-$\tfrac{3}{2}$ model. The results for $D=8,10,12$ consistently show the longest correlation length in the fermionic sector, whereas at $D=6$ the fermionic sector instead exhibits the shortest correlation length. This qualitative discrepancy indicates that the $D=6$ data point is not yet in the same convergence regime and is therefore excluded from the extrapolation.

Second, fitting with a leading-order ansatz ($\sim 1/D^\alpha$) becomes less reliable when small-$D$ data are included, as subleading corrections become non-negligible in that regime. In our calculations, we observe that the $S=\tfrac{3}{2}$ and $S=2$ models require larger bond dimensions to achieve a level of energy convergence comparable to the $S=1$ case. This further motivates the exclusion of smaller-$D$ data points from the extrapolation for these higher-spin systems.

\begin{figure}[!ht]
    \centering
    \includegraphics[width=0.4\textwidth]{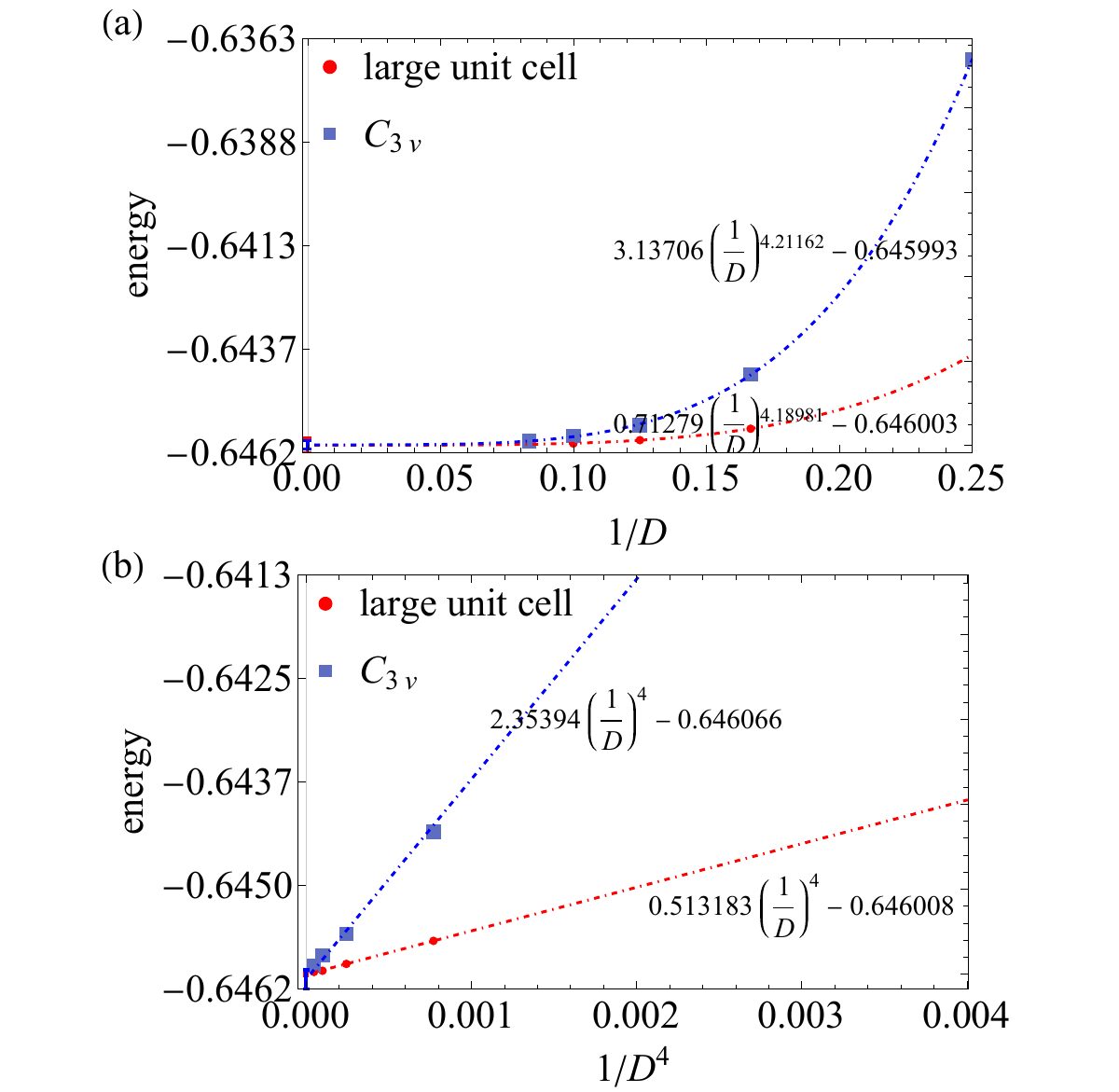}
    \caption{(a) Extrapolation of the energy with $1/D^{\alpha}$ scaling for both our symmetry-breaking iPEPS (large unit cell) and symmetry-restricted iPEPS results. (b) Heuristic choice for $\alpha=4$.}
    \label{fig: C_3v_S1_extrapolation}
\end{figure}

We now compare the extrapolated energy for the spin-1 model obtained by both our symmetry-breaking iPEPS (large unit cell) and symmetry-restricted iPEPS (\cref{fig: C_3v_S1_extrapolation}). The extrapolated results [\cref{fig: C_3v_S1_extrapolation}(a)] improve the agreement from $1\times 10^{-4}$ (at largest $D$) to $1\times 10^{-5}$ (extrapolated value) and find agreement within the error bars.

\section{Additional iPEPS correlation length data}\label{sec:FCLS}
\begin{figure}[!ht]
    \centering
    \includegraphics[width=0.45\textwidth]{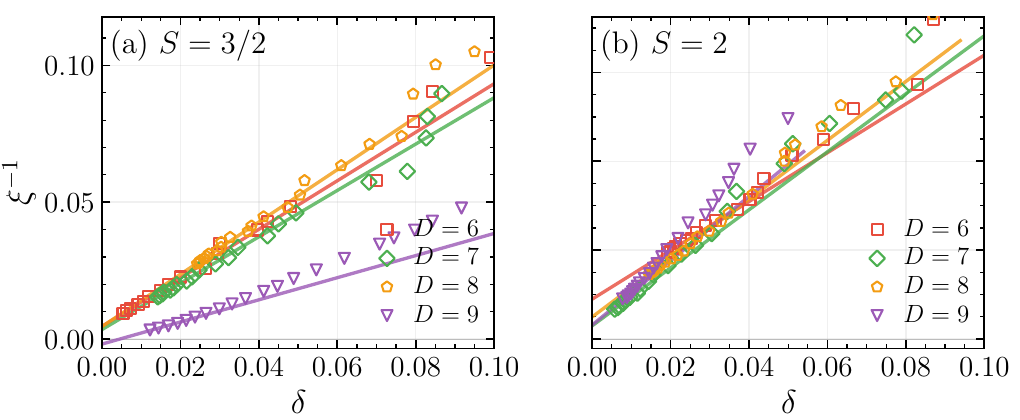}
    \caption{Additional data of finite correlation length scaling for (a) spin-$\tfrac{3}{2}$ and (b) spin-2. $D$ is the bond dimension of PEPS, while the scaling parameter is $\delta=\ln(|\lambda_1/\lambda_3|)$ with $\lambda_i$ being the $i$th largest-magnitude eigenvalue of transfer matrices and the inverse dominant correlation length is given by $\xi^{-1}=-\ln(|\lambda_1/\lambda_0|)$.}
    \label{fig:FCLS-S32}
\end{figure}

In Figs.~\ref{fig:dimer_correlation_FCLS}(b) and \ref{fig:dimer_correlation_FCLS}(c) of the main text, we present the iPEPS data for the correlation length of the spin-$\tfrac{1}{2}$ and spin-1 models. We observe that the data of the spin-$\tfrac{1}{2}$ model show a collapse for different $D$, whereas this feature is absent for the spin-1 model. In \cref{fig:FCLS-S32} we show that the feature is also absent for the spin-$\tfrac{3}{2}$ and spin-2 models.

\section{Solutions of isolated plaquette}\label{sec:plaquette}
In this section, we discuss the solution of the Kitaev model on a single plaquette (six sites). As we discuss in the main text, when the ground-state solution is unique, the prototype trivial gapped ground state with plaquette order can be directly constructed. We find that the ground state of the spin-1 plaquette is unique with energy $\approx -3.63$. The gap to the excited state is $\approx 0.18$. For spin-$\tfrac{1}{2}$ and spin-$\tfrac{3}{2}$, the ground state is, by contrast, fourfold degenerate.

\section{Evidence for plaquette order based on finite-size calculations}\label{sec:finitesize}
In this section, we focus on the spin-1 Kitaev model and present evidence for plaquette order based on high-precision finite-size data.

\begin{figure}[!ht]
    \centering
    \includegraphics[width=0.5\textwidth]{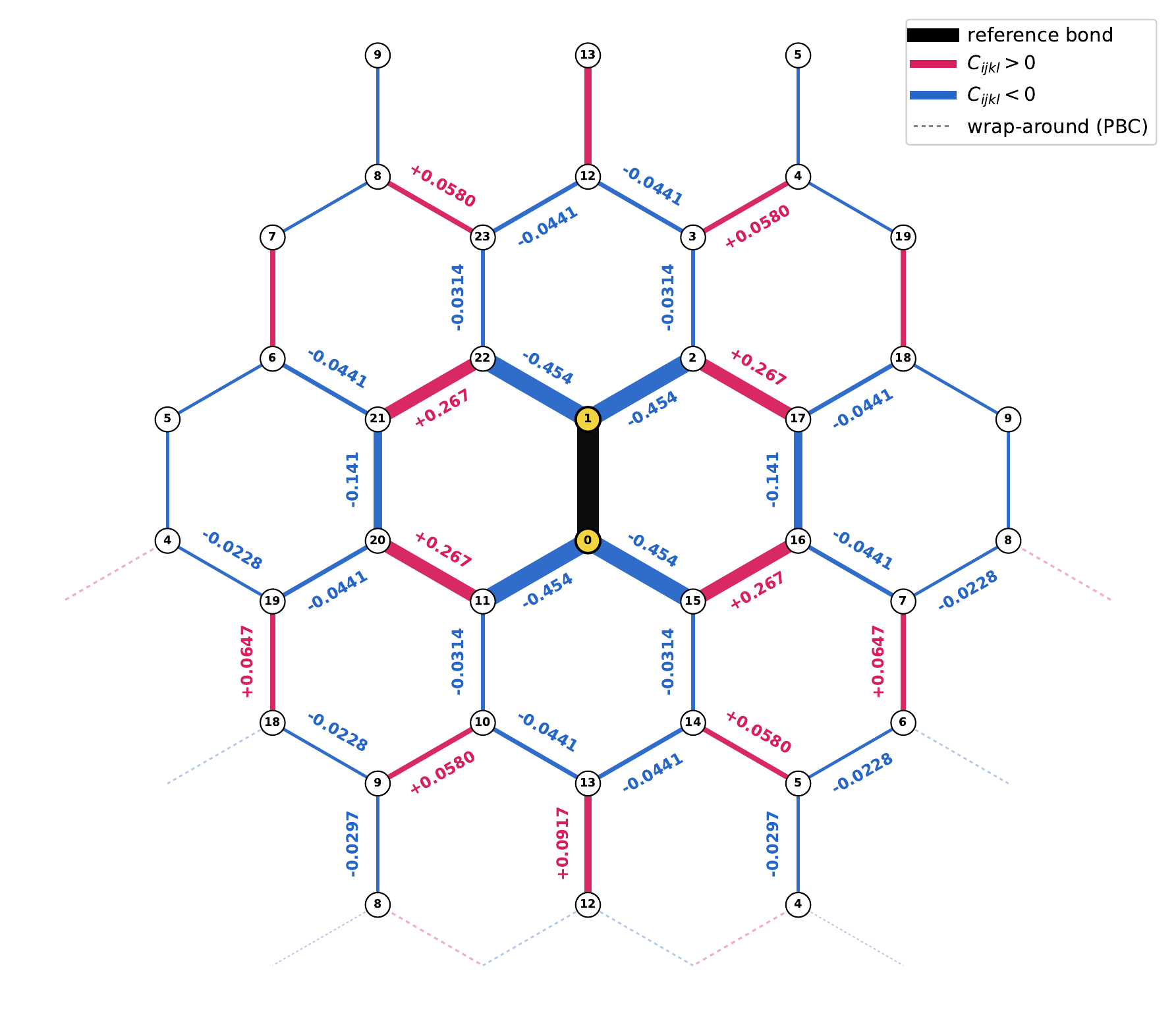}
    \caption{Connected bond-bond correlation function for spin-1 Kitaev model on 24-site hexagonal geometry. The plotted values are $4C_{D_{ij},D_{kl}}=4(\left\langle S^{\gamma_{ij}}_i S^{\gamma_{ij}}_j S^{\gamma_{kl}}_k S^{\gamma_{kl}}_l \right\rangle-\left\langle S^{\gamma_{ij}}_i S^{\gamma_{ij}}_j \right\rangle \left\langle S^{\gamma_{kl}}_k S^{\gamma_{kl}}_l \right\rangle)$.}
    \label{fig: dimer_correlation}
\end{figure}

\begin{figure}[!ht]
    \centering
    \includegraphics[width=0.3\textwidth]{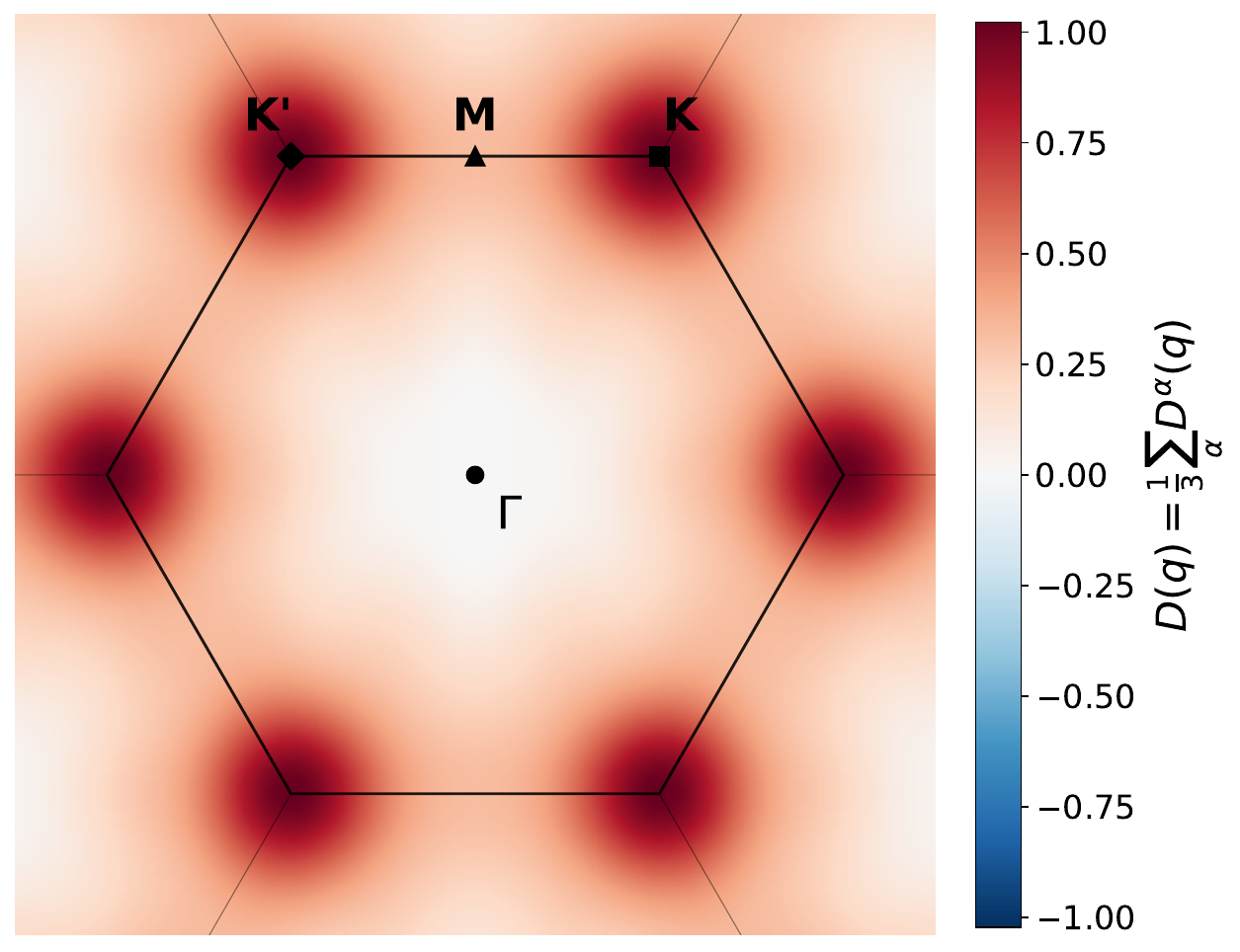}
    \caption{Structure factor of bond-bond correlation function. The real-space bond-bond correlation function (\cref{fig: dimer_correlation}) is Fourier transformed to momentum space with an extended momentum mesh.}
    \label{fig: dimer_structural}
\end{figure}

Using exact diagonalization, we study the model on a 24-site hexagonal cluster with periodic boundary conditions. This geometry preserves the full spatial symmetry of the model and is compatible with plaquette valence-bond-solid (VBS) order. We verify that the ground-state energy is consistent with previous results reported in Ref.~\onlinecite{JPSJ.87.063703}. To probe possible VBS ordering, we compute the bond-bond correlation function~\cite{PhysRevB.84.024406} and its corresponding structure factor. In \cref{fig: dimer_correlation}, we present the connected bond-bond correlations, which exhibit pronounced plaquette correlations. This behavior is similar to that observed in the $J_1$--$J_2$--$J_3$ Heisenberg model on the honeycomb lattice~\cite{PhysRevB.84.024406}. In \cref{fig: dimer_structural}, we show the structure factor of the averaged bond-bond correlations. A clear peak appears at the $K$ and $K'$ points, indicating that the 24-site cluster has a strong susceptibility to the tripled-unit-cell valence-bond order.

\begin{figure}[!ht]
    \centering
    \includegraphics[width=0.5\textwidth]{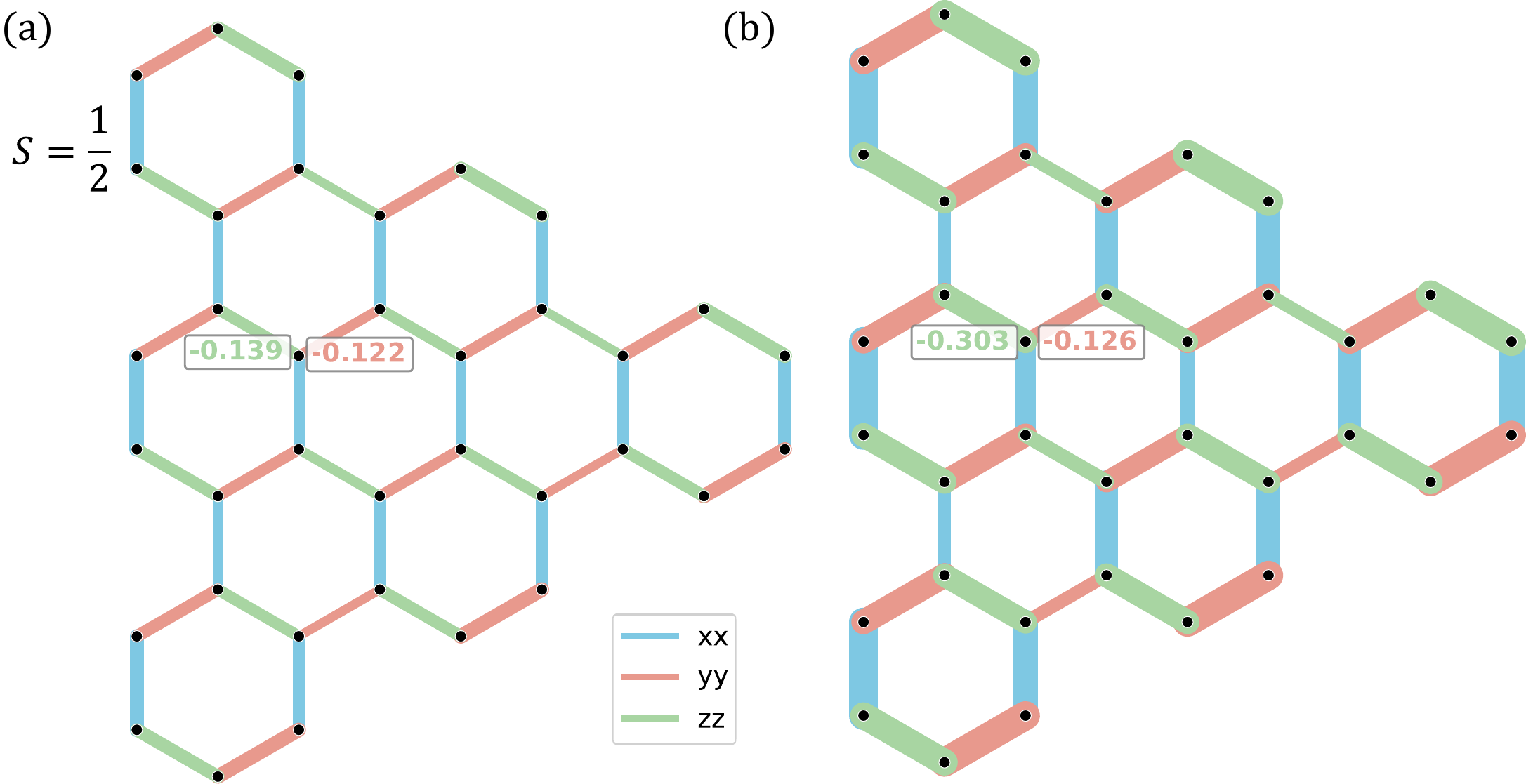}
    \caption{Bond energies for (a) spin-$\tfrac{1}{2}$ and (b) spin-1 Kitaev models on a finite triangle flake. The thickness of a bond is proportional to the absolute value of bond energy. The values of two bonds are denoted in the figures. The data are obtained from DMRG calculations. The bond dimension that we use is 4800 to ensure sufficient precision.}
    \label{fig: Kitaev_Traingle}
\end{figure}

Based on DMRG calculations, we further employ a geometry-induced edge-pinning approach to probe plaquette order~\cite{PhysRevLett.110.127203}. As illustrated in \cref{fig: Kitaev_Traingle}, the honeycomb lattice is arranged on a triangular flake. At the boundary, missing bonds leave spins with two remaining bonds, which can preferentially form stronger plaquette bonds. This geometry, therefore, tends to pin a plaquette with six strong bonds near the edge. Assuming plaquette order, the central plaquette is expected to exhibit alternating strong and weak bonds in this setup.

We estimate the plaquette bond ratio using the bond energies labeled in \cref{fig: Kitaev_Traingle}. For comparison, we also analyze the spin-$\tfrac{1}{2}$ Kitaev model on the same geometry. For the spin-1 case, the estimated bond ratio is approximately $-0.303 / -0.126 \approx 2.4$, which likely overestimates the true value due to finite-size and boundary effects. By contrast, the spin-$\tfrac{1}{2}$ model yields nearly uniform bond energies, consistent with the absence of long-range valence-bond order. In principle, a more accurate bond ratio could be obtained by extrapolating with increasing flake size. However, this is not feasible with DMRG, as the computational cost grows exponentially with the system width and higher spin values require significantly larger bond dimensions. As discussed in the main text, our iPEPS calculations provide a more reliable estimate of the bond ratio.

\end{document}